\documentclass[aps,showpacs,twocolumn]{revtex4}
\usepackage{amsmath}
\usepackage{epsfig}

\begin{document}

\title{Finite axisymmetric charged dust disks in conformastatic spacetimes}

\author{Guillermo A. Gonz\'{a}lez}\email[Email address: ]{guillego@uis.edu.co}
\affiliation{Escuela de F\'{\i}sica, Universidad Industrial de Santander, A. A.
678, Bucaramanga, Colombia}
\affiliation{Departamento de F\'{\i}sica Te\'orica, Universidad del Pa\'is
Vasco, 48080 Bilbao, Spain}

\author{Antonio C. Guti\'errez-Pi\~{n}eres}\email[Email address: ]
{gutierrezpac@yahoo.com}
\author{Paolo A. Ospina}\email[Email address: ]{paolo6506@gmail.com}
\affiliation{Escuela de F\'{\i}sica, Universidad Industrial de Santander, A. A.
 678, Bucaramanga, Colombia}


\begin{abstract}

An infinite family of axisymmetric charged dust disks of finite extension is
presented. The disks are obtained by solving the vacuum Einstein-Maxwell
equations for conformastatic spacetimes, which are characterized by only one
metric function. In order to obtain the solutions, it is assumed that the metric
function and the electric potential are functionally related and that the metric
function is functionally dependent of another auxiliary function, which is taken
as a solution of Laplace equation. The solutions for the auxiliary function are
then taken as given by the infinite family of generalized Kalnajs disks recently
obtained by Gonz\'alez and Reina \cite{GR}, which is expressed in terms of the
oblate spheroidal coordinates and represents a well behaved family of finite
axisymmetric flat galaxy models. The so obtained relativistic thin disks have
then a charge density that is equal, except maybe by a sign, to their mass
density, in such a way that the electric and gravitational forces are in exact
balance. The energy density of the disks is everywhere positive and well
behaved, vanishing at the edge. Accordingly, as the disks are made of dust,
their energy-momentum tensor it agrees with all the energy conditions.

\end{abstract}

\maketitle

\section{Introduction}

The study of axially symmetric solutions of Einstein and Einstein-Maxwell field
equations corresponding to disklike configurations of matter, apart from its
purely mathematical interest, has a clear astrophysical relevance. Indeed, thin
disks can be used to model accretion disks, galaxies in thermodynamical
equilibrium and the superposition of a black hole and a galaxy. Disk sources for
stationary axially symmetric spacetimes with  magnetic fields are also of
astrophysical importance  mainly in the study of neutron stars, white dwarfs and
galaxy formation. Now, although normally it is considered that disks with
electric fields do not have clear astrophysical importance, there exists the
posibility that some galaxies be positively charged \cite{BH}, so that the study
of charged disks may be  of interest not only in the context of exact solutions.
Accordingly, through the years, many attempts had been made to find exact
solutions, static and stationary, to the Einstein and Einstein-Maxwell equations
that have as its source a relativistic thin disk. 

Exact solutions having as sources relativistic static thin disks were first
studied by Bonnor and Sackfield \cite{BS} and Morgan and Morgan \cite{MM1,MM2}.
Since then several classes of exact solutions corresponding to static \cite{VOO,
LP1, CHGS, LO, LEM, BLK, BLP, GL1, GE} and stationary \cite{LP2, BL, PL, GL2}
thin disks have been obtained by different authors and the superposition of a
static or stationary thin disk with a black hole has been considered in 
\cite{LL1, LL2, LL3, SZ1, SEM1, SZ2, SEM2, SEM3, KHZ}. Relativistic disks
embedded in an expanding FRW universe have been studied in \cite{FIL}, perfect
fluid disks with halos in \cite{VL1} and the stability of thin disks models has
been investigated using a first order perturbation of the energy-momentum tensor
in \cite{UL1}. On the other hand, thin disks have been discussed as sources for
Kerr-Newman fields \cite{LBZ,GG1}, magnetostatic axisymmetric fields \cite{LET1}
and conformastatic and conformastationary metrics \cite{VL2,KBL}, while models
of electrovacuum static counterrotating dust disks were presented in \cite{GG2}.
Charged perfect fluid disks were also studied in \cite{VL3}, and  charged
perfect fluid disks as sources of  static and  Taub-NUT-type spacetimes in
\cite{GG3,GG4}.

Now, between the above mentioned works, particularly interesting are those that
consider dust disks in comformastatic spacetimes \cite{KBL,VL2}. In this case,
the charge density of the disks is equal to their mass density and so the
electric and gravitational forces are in exact balance. This kind of equilibrium
configuration has been called by some authors `electrically counterpoised dust'
(ECD) and has been studied with some detail, both in classical and relativistic
theories \cite{B,BB,BW,GUR,VAR}. Now, as the matter content of the source is
dust, the corresponding energy-momentum tensor very probably will agree with all
the energy conditions, a fact that is particularly relevant for models of
relativistic thin disks. Indeed, as can be see in some of the above mentioned
works, there are many models of relativistic thin disk that do not agree with
these conditions, and also many models that only fulfill these conditions
partially.

The conformastatic thin disks presented at references \cite{KBL,VL2} were
obtained by means of the well known `displace, cut and reflect' method in order
to introduce a discontinuity at the first derivative of one otherwise smooth
solution. The result is a solution with a singularity of the delta function type
in all the $z = 0$ hypersurface and so can be interpreted as an infinite thin
disk. On the other hand, solutions that can be interpreted as thin disks of
finite extension can be obtained if a proper coordinate system is introduced. A
coordinate system that it adapts naturally to a finite source and it presents
the required discontinuous behavior is given by the oblate spheroidal
coordinates. Some examples of finite thin disks solutions expressed in these
coordinates can be found at references \cite{BS,MM1,VOO,LO}.

In this paper we present an infinite family of axially symmetric charged dust
disks of finite extension. In order to obtain the thin disk solutions, we will
solve the Einstein-Maxwell equations for conformastatic spacetimes assuming that
the metric function and the electric potential are functionally related and that
the metric function is functionally dependent of another auxiliary function,
which is taken as a solution of Laplace equation. Then we take the solutions for
the auxiliary potential as given by the infinite family of generalized Kalnajs
disks recently obtained by Gonz\'alez and Reina \cite{GR}, which is expressed in
terms of the oblate spheroidal coordinates and represents a well behaved family
of finite axisymmetric flat galaxy models.

Accordingly, the paper is organized as follows. First, in Sec. \ref{sec:ecs}, we
present the vacuum Einstein-Maxwell equations system and their solution for
conformastatic spacetimes. We introduce the assumed functional dependences in
order to explicitly integrate the equations system in such a way that the metric
function and the electric potential can be expressed in terms of solutions of
Laplace equation. The oblate spheroidal coordinates are introduced and the
general solution of the Laplace equation expressed in these coordinates is
presented.

Next, in Sec. \ref{sec:emt}, we obtain the surface energy-momentum tensor and
the surface current density of the relativistic thin disks. So, we first present
a summary of the procedure to obtain the energy-momentum tensor and the current
density by using the distributional approach for the case of conformastatic
spacetimes. The relation between the energy density and the charge density is
explicitly derived and the energy density is written in terms of the mass
density of the Newtonian thin disks, in such a way that is explicitly shown that
the relativistic thin disks will be in agreement with all the energy conditions.

Then, in Sec. \ref{sec:kal}, we restrict the general model by considering a
particular family of well behaved charged dust disks. We present the particular
form of the solutions of Laplace equation that describe finite Newtonian thin
disk with a well behaved surface mass density, and we use this solution to
obtain the corresponding relativistic charged dust disks. The behavior of the
energy density is then graphically analyzed and their main properties are
described. Finally, in Sec. \ref{sec:conc}, we present a brief discussion of our
main results.

\section{\label{sec:ecs}Einstein-Maxwell equations and finite thin disk
solutions}

The vacuum Einstein-Maxwell equations, in geo\-me\-trized units such that $c = G
= \mu _{0} = \epsilon _{0} =  1$, can be written as 
\begin{eqnarray}
G_{ab} &=& 8 \pi \ T_{ab},  \label{eq:emep1} \\
&   &     \nonumber    \\
{F^{ab}}_{;b} &=& 0, \label{eq:emep2}
\end{eqnarray}
with the electromagnetic energy-mementum tensor given by
\begin{equation}
T_{ab} = \frac{1}{4 \pi} \left[ F_{ac} F_b^{ \ c} - \frac{1}{4} g_{ab} F_{cd}
F^{cd} \right], \label{eq:emtensor}  
\end{equation}
where
\begin{equation}
F_{ab} =  A_{b,a} - A_{a,b} \label{eq:fab}
\end{equation}
is the electromagnetic field tensor and $A_a$ is the electromagnetic four
potential.

Now then, for a conformastatic spacetime the line element can be written in
cylindrical coordinates $x^a = (t, \varphi, r, z)$ as \cite{SYN}
\begin{equation}
ds^2 = - \ e^{2\lambda} dt^2  +  e^{- 2\lambda} (r^2 d\varphi^2 + dr^2 + dz^2),
\label{eq:metCC} 
\end{equation} 
where the metric function $\lambda$ do not depends on $t$. So, if we take the
electromagnetic potential as
\begin{equation}
A_a = (- \phi, 0, 0, 0),
\end{equation}
where it is assumed that the electric potential $\phi$ also is independent of
$t$, the vacuum Einstein-Maxwell equations reduce to
\begin{eqnarray}
\nabla^2 \lambda &=& e^{- 2 \lambda} \nabla \phi \cdot \nabla \phi,
\label{eq:eme2} \\
&  & \nonumber \\
\nabla^2 \phi &=& 2 \ \nabla \lambda \cdot \nabla \phi, \label{eq:eme3}\\
 &  & \nonumber \\
\lambda_{,i} \lambda_{,j} &=& e^{- 2 \lambda} \phi_{,i} \ \phi_{,j},
\label{eq:eme1} 
\end{eqnarray}
where $i,j = 1,2,3$ and $\nabla$ is the usual differential operator in
cylindrical coordinates.

In order to solve the above equations system, first it is assumed that the
electric potential $\phi$ is functionally dependent of the metric function
$\lambda$, $\phi = \phi(\lambda)$, so that equation (\ref{eq:eme1}) implies that
\begin{eqnarray}
[\phi'(\lambda)]^{2}=e^{2\lambda},
\end{eqnarray}
whose solution is given by
\begin{equation}
\phi = \pm e^{\lambda} + k_1, \label{eq:philam} 
\end{equation}
with $k_1$ an arbitrary integration constant. With this solution, the equations
system (\ref{eq:eme2})-(\ref{eq:eme3}) reduces to only one non-linear partial
differential equation,
\begin{equation}
\nabla^{2} \lambda = \nabla \lambda \cdot \nabla \lambda, \label{eq:eclam}
\end{equation}
for the metric function $\lambda$. Then, we assume an additional functional
dependence for the metric function  $\lambda$, which it is expressed as
$\lambda= \lambda ( U  )$, where $U$ is an auxiliary function that is taken as a
solution of Laplace equation. Thus then, equation (\ref{eq:eclam}) it reduces to
\begin{equation}
\lambda''(U) = \left[ \lambda'(U) \right]^{2}, \label{12} 
\end{equation}
whose solution is given by
\begin{equation}
e^{\lambda} = \frac{k_{3}}{U + k_{2}}, \label{lambdau} 
\end{equation}
where $k_2$ and $k_3$ are arbitrary integration constants. 

Now, in order to have an appropriated behavior at infinity, we will impose some
boundary conditions on the solutions. So, in order to have an assymptotically
flat spacetime, we must have that $e^\lambda = 1$ at infinite. Also we require
that $\phi = 0$ at infinity, since we are only considering sources of finite
extension. Accordingly, we shall consider for the function $U$ only those
solutions of Laplace equation that behave as $U = 0$ at infinite, in such a way
that by taking $k_2 = k_3$ we obtain the desired behavior for $\lambda$.
Finally, in order to fulfill the required behavior for $\phi$, we must take $k_1
= \mp 1$ and so equation (\ref{eq:philam}) can be cast as
\begin{equation}
\phi = \pm \left[ \frac{k}{U + k} - 1 \right], \label{eq:phiu}
\end{equation}
where the constants have been renamed as $k_2 = k_3 = k$. 

As we can see from (\ref{lambdau}) and (\ref{eq:phiu}), the solutions of the
Einstein-Maxwell equations system (\ref{eq:eme2})-(\ref{eq:eme3}) are expressed
in terms of $U$, a solution of the Laplace equation. So, in order to have
solutions that correspond to finite thin disks we must to use solutions of
Laplace equation that properly describe Newtonian disklike sources of finite
radius. This kind of solutions can be obtained by introducing the oblate
spheroidal coordinates, which adapt in a natural way to the geometry of these
sources. This coordinates are related with the cylindrical coordinates by the
relation \cite{MF}
\begin{eqnarray} 
r^{2} &=& a^{2}(1 + \xi^{2})(1 - \eta^{2}), \label{eq:ciloblatas1} \\
& & 	\nonumber \\
z &=& a \xi \eta, \label{eq:ciloblatas2}
\end{eqnarray}
where $0 \leq \xi < \infty$ and $-1 \leq \eta < 1$. So, the disk has coordinates
$\xi = 0$, $0\leq\eta^2<1$ and, on crossing the disk, $\eta$ changes sign but
does not change in absolute value.

Now then, the singular behavior of the coordinate $\eta$ implies that a
polynomial in even powers of $\eta$ is a continuous function everywhere but has
a discontinuous $\eta$-derivative at the disk. Accordingly, the general solution
of Laplace equation corresponding to an axially symmetric thin disklike source
of radius $a$ can be expressed as \cite{BAT}
\begin{equation}
U(\xi,\eta) = - \sum_{n=0}^{\infty} C_{2n} q_{2n}(\xi) P_{2n}(\eta),
\label{eq:gensol}
\end{equation}
where the $P_{2n}(\eta)$ are the legendre polynomials of order $2n$ and
$q_{2n}(\xi) = i^{2n+1} Q_{2n}(i\xi)$, with $Q_{2n}(i\xi)$ the legendre function
of second kind of imaginary argument. The $C_{2n}$ are constants that must be
specified in order to have any particular solution and are directly related with
the surface mass density of the Newtonian source considered. So, in a next
section, we will present a particular choice of these constants corresponding to
an infinite family of finite thin disks with a well behaved surface mass
density, the family of Kalnajs generalized disks recently presented by
Gonz\'alez and Reina \cite{GR}.

\section{\label{sec:emt}Energy-momentum tensor and current density} 

As was pointed in the precedent section, the solutions of the Einstein-Maxwell
equations corresponding to a finite disklike source are even functions of the
$z$ coordinate. So then, they are everywhere continuous functions but with their
first $z$-derivatives discontinuous at the disk surface. Accordingly, in order
to obtain the energy-momentun tensor and the current density of the source, we
will express the jump across the disk of the first $z$-derivatives of the metric
tensor as 
\begin{equation}
 b_{ab} =  [{g_{ab,z}}] = 2 {g_{ab,z}}|_{_{z = 0^+}},
\end{equation}
and the jump across the disk of the electromagnetic field tensor as
\begin{equation}
[F_{za}] = [A_{a,z}] = 2 {A_{a,z}}|_{_{z = 0^+}},
\end{equation}
where the reflection symmetry of the functions with respect to $z = 0$ has
been used.

Then, by using the distributional approach \cite{PH, LICH, TAUB}, the Einstein-Maxwell equations yield an energy-momentum tensor as
\begin{equation}
T^{ab} = T_+^{ab} \theta(z) + T_-^{ab} [1 - \theta(z)] + Q^{ab} \delta(z),
\label{eq:emtot}
\end{equation}
and a current density as
\begin{equation}
J^a = I^a \delta(z),   \label{eq:courrent}
\end{equation}
where $\theta(z)$ and $\delta (z)$ are, respectively, the Heaveside and Dirac distributions with support on $z = 0$. Here $T_\pm^{ab}$ are the electromagnetic energy-momentum tensors as defined by (\ref{eq:emtensor}) for the $z \geq 0$ and $z \leq 0$ regions, respectively, whereas that
\begin{eqnarray}
16 \pi Q^a_b &=& b^{az}\delta^z_b - b^{zz}\delta^a_b + g^{az}b^z_b -
g^{zz}b^a_b \nonumber \\
&& \ + \ b^c_c (g^{zz}\delta^a_b - g^{az}\delta^z_b)
\end{eqnarray}
gives the part of the energy-momentum tensor corresponding to the disk source, and 
\begin{equation}
4 \pi I^a  =[F^{az}]
\end{equation}
is the contribution of the disk source to the current density. Now, the ``true'' surface energy-momentum tensor of the disk, $S_{ab}$, and the
``true'' surface current density, $j^a$, can be obtained through the
relationships
\begin{eqnarray}
S_{ab} &=& \int Q_{ab} \ \delta (z) \ ds_n \ = \ e^{ - \lambda} \ Q_{ab} \ ,    \\
&   &           \nonumber      \\
j^a &=& \int I^a \ \delta (z) \ ds_n \ = \  e^{- \lambda} I^a ,
\end{eqnarray}
where $ds_n = \sqrt{g_{zz}} \ dz$ is the physical measurement of length in
the direction normal to the disk. 

For the metric (\ref{eq:metCC}), the only non-zero component of $Q^a_b$ is
\begin{equation}
Q^0_0 = - \frac{e^{2\lambda} \lambda_{,z}}{2 \pi}, \label{eq:emt}
\end{equation}
whereas that the only non-zero component of $I^a$ is
\begin{equation}
I^0 = - \frac{\phi_{,z}}{2 \pi}.
\end{equation}
Thus, the only non-zero component of surface energy-momentum tensor $S^a_b$ is given by 
\begin{equation}
S^0_0 = - \frac{e^{\lambda} \lambda_{,z}}{2 \pi}, \label{eq:emts}
\end{equation}
and the only non-zero component of the surface current density $j^a$ is
\begin{equation}
j^0 = \ - \frac{e^{- \lambda} \phi_{,z}}{2 \pi},  \label{eq:37}
\end{equation}
where all the quantities are evaluated at $z = 0^+$.

The surface energy-momentum tensor of the disk and the surface current density
of the disk then can be written as
\begin{eqnarray}
S^{ab} &=& \epsilon V^a V^b , \\
 &	&	\nonumber	\\
j^a &=& \sigma V^a ,
\end{eqnarray}
where
\begin{equation}
V^a = e^{-\lambda} (1, 0, 0, 0 ) 
\end{equation}
is the velocity vector of the matter distribution. So then, the energy
density and the charge density of the distribution of matter are given by
\begin{eqnarray}
\epsilon &=&  \frac{e^\lambda \lambda_{,z}}{2 \pi} , \label{eq:epsi} \\
 & & 	\nonumber	\\
\sigma &=&  - \frac{\phi_{,z}}{2 \pi} \label{eq:sigma} ,
\end{eqnarray}
respectively. Now, by using equation (\ref{eq:philam}), the expression
(\ref{eq:sigma}) can be written as
\begin{equation}
\sigma = \mp \ \epsilon  \label{eq:sigma*}
\end{equation}
so that the charge density of the disks is equal, except maybe by a sign, to
their mass density. Accordingly, the electric and gravitational forces are in
exact balance, as in the configurations of ECD that were mentioned at the
introduction.

Now then, in the context of classical general relativity, it is assumed that the
energy-momentum tensor must fulfill certain requirements, which are embodied in
the {\it weak, strong and dominat energy conditions} \cite{HE}. Indeed, for the
case of a dust source, all these conditions it reduce to the condition that the
energy density be greater or equal to zero, $\epsilon \geq 0$. On the other
hand, from equation (\ref{lambdau}), we have that the energy density can be
written as
\begin{equation}
\epsilon = - \frac{k \Sigma}{(U + k)^2} , \label{eq:epsig}
\end{equation}
where
\begin{equation}
\Sigma = \frac{U_{,z}}{2 \pi}
\end{equation}
is the Newtonian mass density of a disklike source which gravitational potential
is given by $U$. Accordingly, if the Newtonian mass density $\Sigma$ is
everywhere no negative, the corresponding relativistic energy density $\epsilon$
will be no negative everywhere only if we take $k < 0$. So then, in order that
the energy-momentum tensor of the disks will agree with all the energy
conditions, the $C_{2n}$ constants in (\ref{eq:gensol}) must be properly chosen
in such a way that $\Sigma \geq 0$. Furthermore, if we have a Newtonian
potential $U$ that it is negative everywhere, as is expected for a compact
Newtonian source, then the energy density $\epsilon$ will be non-singular
everywhere at the disk.

\section{\label{sec:kal}A particular family of disks}

Now we shall restrict the previous general model by considering a particular
family of disks with a well-behaved surface energy density. The members of the
family are expressed in terms of particular solutions $U_m$ of the Laplace
equation obtained by choosing properly the constants of the general solution
(\ref{eq:gensol}). The obtained particular solutions $U_m$ represent the 
Newtonian gravitational potential of finite thin disks with mass density given
by
\begin{equation}
\Sigma_{m} (r) = \frac{(2m+1)M}{2\pi a^{2}} \left[ 1 - \frac{r^{2}}{a^{2}}
\right]^{m - \frac{1}{2}}, 
\end{equation}
where $M$ and $a$ are, respectively, the total mass and the radius of the disk and we must take $m \geq 1$. For each value of $m$, the constants $C_{2n}$ are defined through the relation \cite{GR}
\begin{equation}
C_{2n} =  \frac{K_{2n}}{(2n+1) q_{2n+1}(0)},
\end{equation}
where
\begin{equation}
K_{2n} = \frac{M}{2a} \left[ \frac{ \pi^{1/2} \ (4n+1) \ (2m+1)!}{2^{2m} (m -
n)! \Gamma(m + n + \frac{3}{2})} \right] 
\end{equation}
for $n \leq m$, and $C_{2n} = 0$ for $n > m$. In Table \ref{tab:c2n} we list the
values of the nonzero $C_{2n}$ for the first six members of the family. So, by
using this constants in the general solution (\ref{eq:gensol}) is easy to see
that the gravitational potential $U_m$ will be negative everywhere, as was
imposed at the previous section.

\begin{table}[t]
\caption{\label{tab:c2n}The $C_{2n}$ constants for $m = 0, ... , 6$.}
\begin{ruledtabular}
\begin{tabular}{lccccccc}
$m$ & ${C}_{0 }$ & ${C}_{2 }$ & ${C}_{4 }$ & ${C}_{6 }$ & ${C}_{8 }$ & ${C}_{10}$ & ${C}_{12}$ \\ \hline
 & & & & & & & \\ 
$1$ & $\frac{M}{a}$& $\frac{M}{a}$ & & & & & \\ 
 & & & & & & & \\ 
$2$ & $\frac{M}{a}$ & $\frac{10 M}{7 a}$ & $\frac{3 M}{7 a}$ & & & & \\ 
 & & & & & & & \\ 
$3$ & $\frac{M}{a}$ & $\frac{5 M}{3 a}$ & $\frac{9 M}{11 a}$ & $\frac{5 M}{33 a}$ & & & \\
 & & & & & & & \\ 
$4$ & $\frac{M}{a}$ & $\frac{20 M}{11 a}$ & $\frac{162 M}{143 a}$ & $\frac{4 M}{11 a}$ &
$\frac{7 M}{143 a}$ & & \\ 
 & & & & & & & \\ 
$5$ & $\frac{M}{a}$ & $\frac{25 M}{13 a}$ & $\frac{18 M}{13 a}$ & $\frac{10 M}{17 a}$ & $\frac{35 M}{247 a}$ & $\frac{63 M}{4199 a}$ & \\ 
 & & & & & & & \\ 
$6$ & $\frac{M}{a}$ & $\frac{2M}{a}$ & $\frac{27 M}{17 a}$ & $\frac{260 M}{323 a}$ & $\frac{5 M}{19 a}$ & $\frac{378 M}{7429 a}$ & $\frac{33 M}{7429 a}$ \\
 & & & & & & & \\ 
\end{tabular}
\end{ruledtabular}
\end{table}

With the above values for the $C_{2n}$, we can easily compute the corresponding energy density of the disks by using equation (\ref{eq:epsig}). Now, in order to graphically illustrate the behavior of the different particular models, first we introduce dimensionless quantities through the relations
\begin{eqnarray}
{\tilde U}_m ({\tilde r}) &=& \frac{a U_m ({\tilde r})}{M}, \\
	&&	\nonumber	\\
{\tilde \Sigma}_m ({\tilde r}) &=& \frac{\pi a^2 \Sigma_m ({\tilde r})}{M}, \\
	&&	\nonumber	\\
{\tilde \epsilon}_m ({\tilde r}) &=& \pi a \epsilon_m ({\tilde r}),
\end{eqnarray}
where ${\tilde r} = r/a$, $0 \leq {\tilde r} \leq 1$, and $U_m ({\tilde r})$ is
evaluated at $z = 0^+$. Accordingly, the dimesionless energy density ${\tilde
\epsilon} ({\tilde r})$ can be written as
\begin{equation}
{\tilde \epsilon}_m ({\tilde r}) = - \frac{{\tilde k} {\tilde \Sigma}_m ({\tilde
r})}{[{\tilde U}_m ({\tilde r}) + {\tilde k}]^2},
\end{equation}
with ${\tilde k} = (k a)/M$.

Then, by using the above expressions and the values of the $C_{2n}$ constants
given at Table \ref{tab:c2n}, we obtain the following expressions
\begin{widetext}\begin{eqnarray}
{\tilde \epsilon}_1 &=& - \frac{3 {\tilde k} \sqrt{1 - {\tilde r}^2}}{2 [{\tilde k} + \frac{3 \pi}{8} ({\tilde r}^2 - 2)]^2}, \\
	&&	\nonumber	\\
{\tilde \epsilon}_2 &=& - \frac{5 {\tilde k} (1 - {\tilde r}^2)^{3/2}}{2 [{\tilde k} - \frac{15 \pi}{128} (3 {\tilde r}^4 - 8 {\tilde r}^2 + 8)]^2}, \\
	&&	\nonumber	\\
{\tilde \epsilon}_3 &=& - \frac{7 {\tilde k} (1 - {\tilde r}^2)^{5/2}}{2 [{\tilde k} + \frac{35 \pi}{512} (5 {\tilde r}^6 - 18 {\tilde r}^4 + 24 {\tilde r}^2 - 16)]^2}, \\
	&&	\nonumber	\\
{\tilde \epsilon}_4 &=& - \frac{9 {\tilde k} (1 - {\tilde r}^2)^{7/2}}{2 [ {\tilde k} - \frac{315 \pi}{32768} (35 {\tilde r}^8 - 160 {\tilde r}^6 + 288 {\tilde r}^4 - 256 {\tilde r}^2 + 128)]^2}, \\
	&&	\nonumber	\\
{\tilde \epsilon}_5 &=& - \frac{11 {\tilde k} (1 - {\tilde r}^2)^{9/2}}{2 [ {\tilde k} + \frac{693 \pi}{131072} (63 {\tilde r}^{10} - 350 {\tilde r}^8 + 800 {\tilde r}^6 - 960 {\tilde r}^4 + 640 {\tilde r}^2 - 256)]^2}, \\
	&&	\nonumber	\\
{\tilde \epsilon}_6 &=& - \frac{13 {\tilde k} (1 - {\tilde r}^2)^{11/2}}{2 [{\tilde k} - \frac{3003 \pi}{2097152} (231 {\tilde r}^{12} - 1512 {\tilde r}^{10} + 4200 {\tilde r}^8 - 6400 {\tilde r}^6 + 5760 {\tilde r}^4 - 3072 {\tilde r}^2 + 1024)]^2},
\end{eqnarray}\end{widetext}
for the energy densities of the first six disk models.

\begin{figure*}
$$\begin{array}{cc}
{\tilde \epsilon}_1  & {\tilde \epsilon}_2  \\
\epsfig{width=2.75in,file=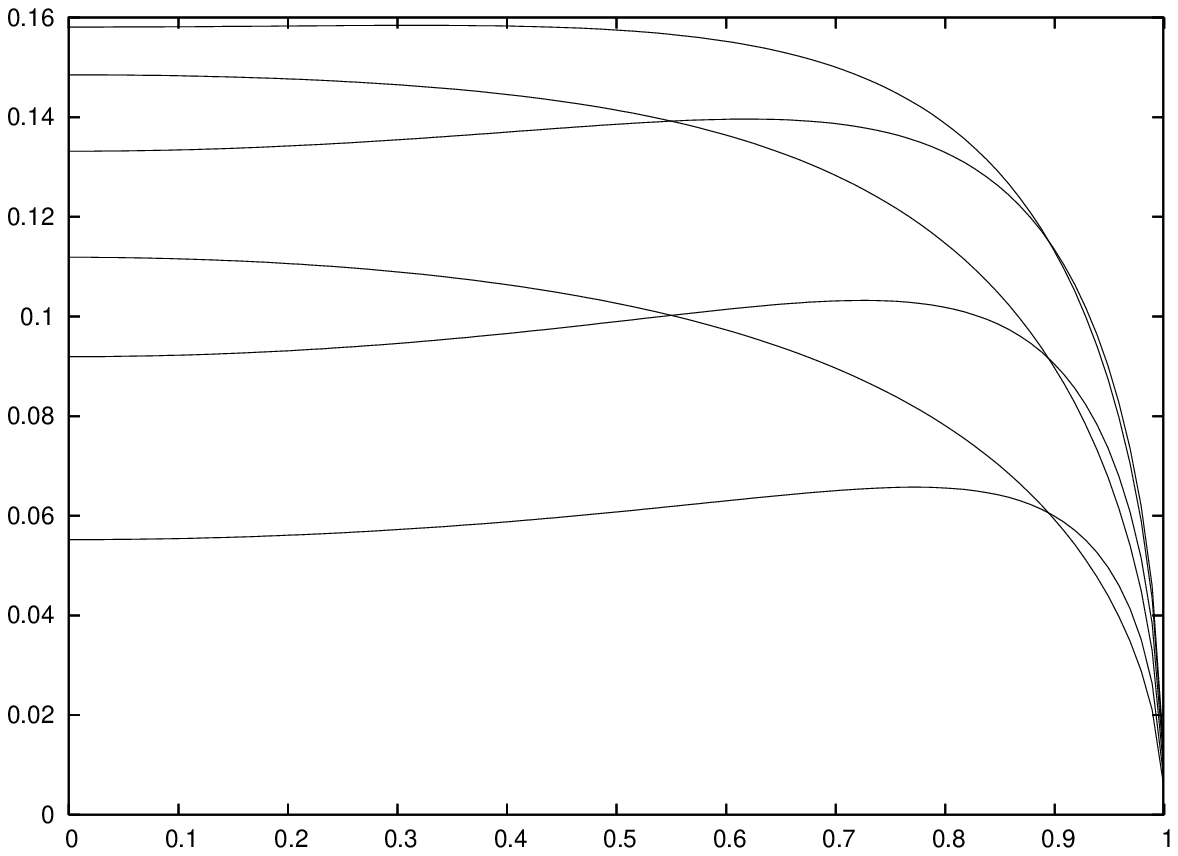} &
\epsfig{width=2.75in,file=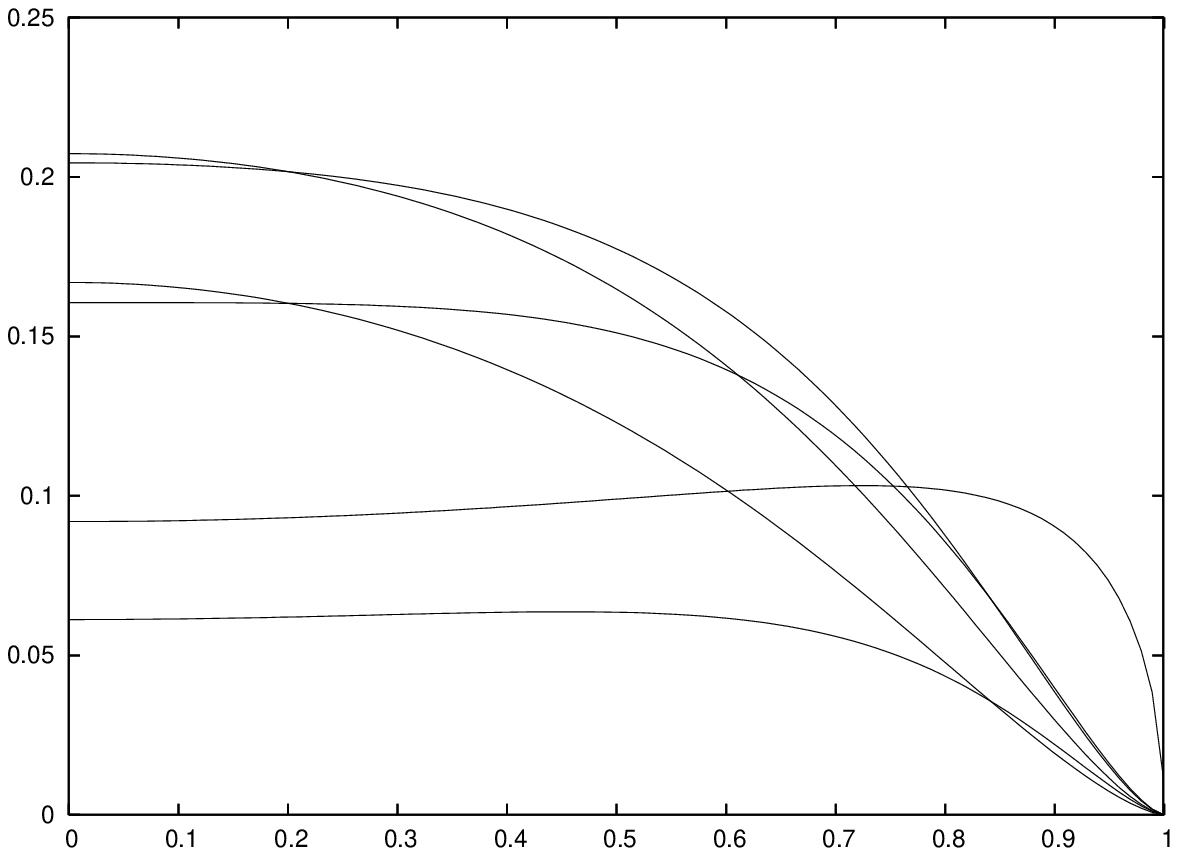} \\
 & \\
{\tilde \epsilon}_3  &  {\tilde \epsilon}_4  \\
\epsfig{width=2.75in,file=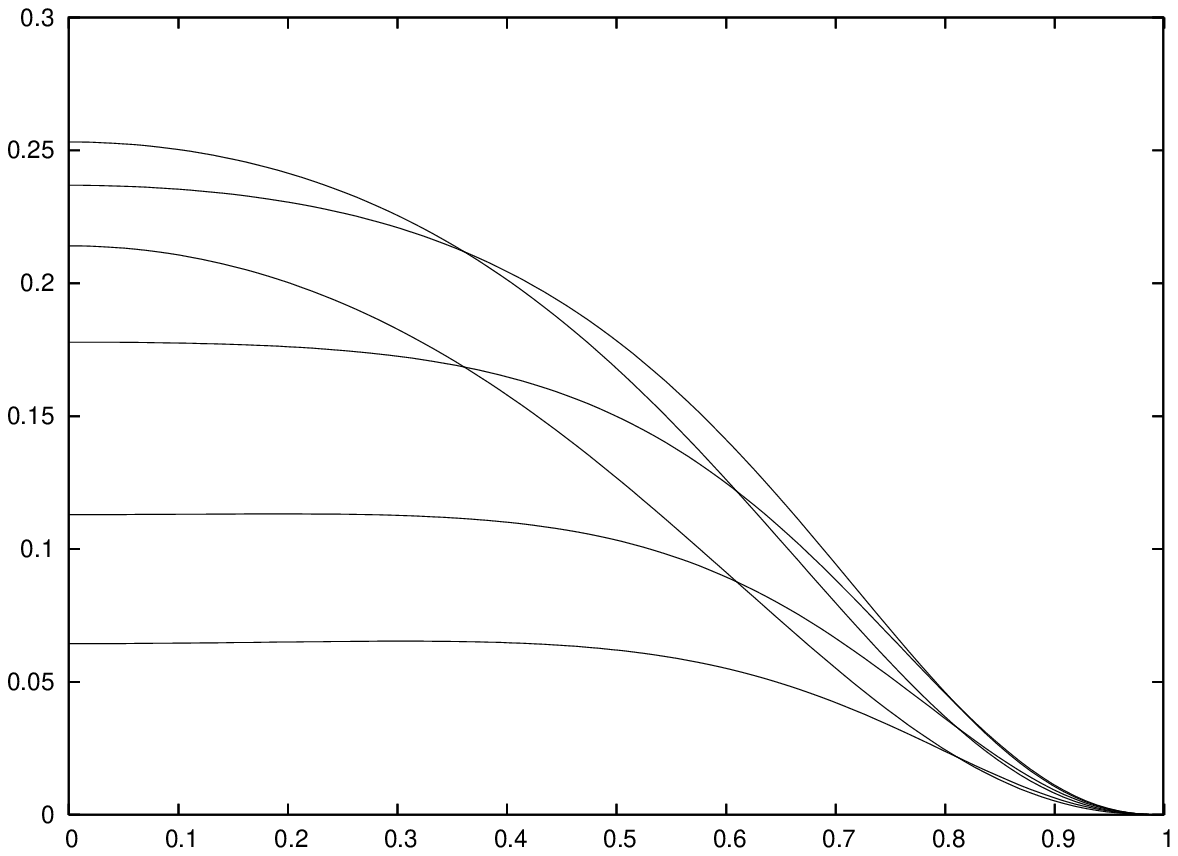} & 
\epsfig{width=2.75in,file=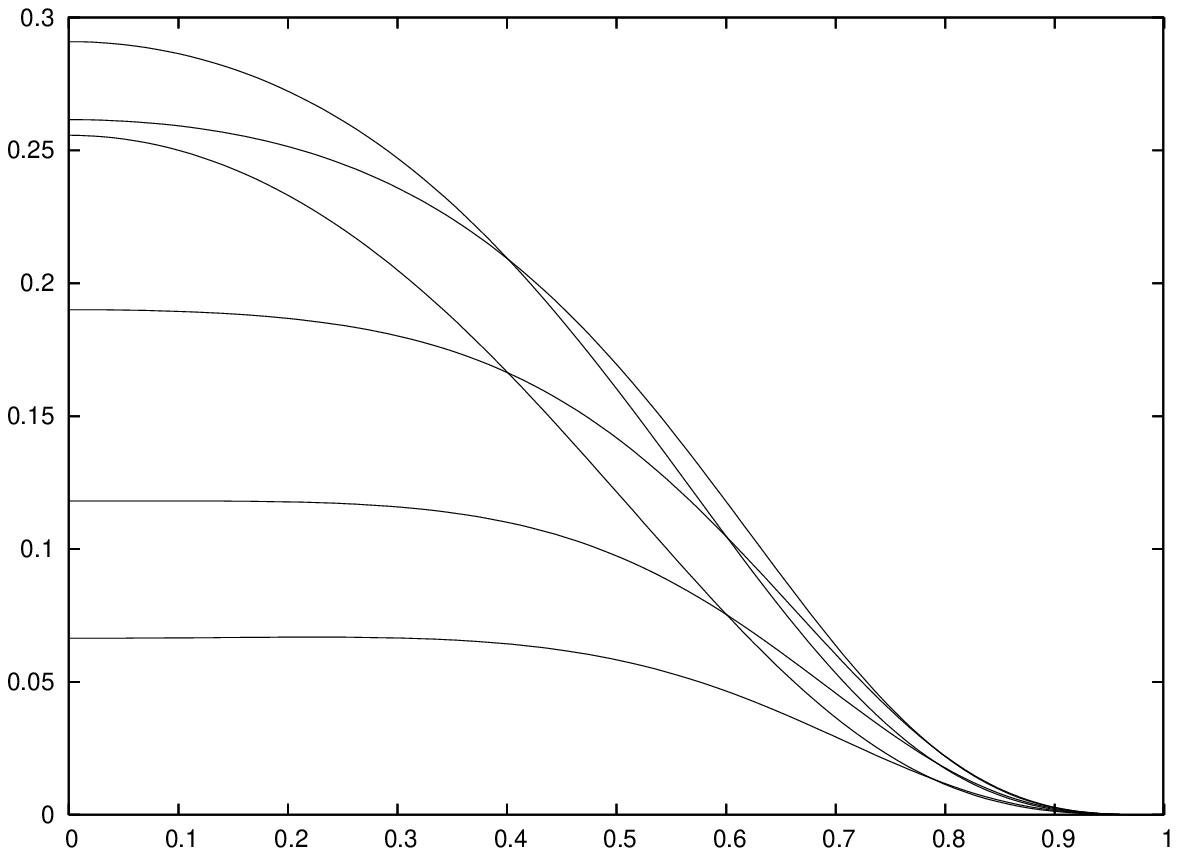} \\
 & \\
{\tilde \epsilon}_5  &  {\tilde \epsilon}_6  \\
\epsfig{width=2.75in,file=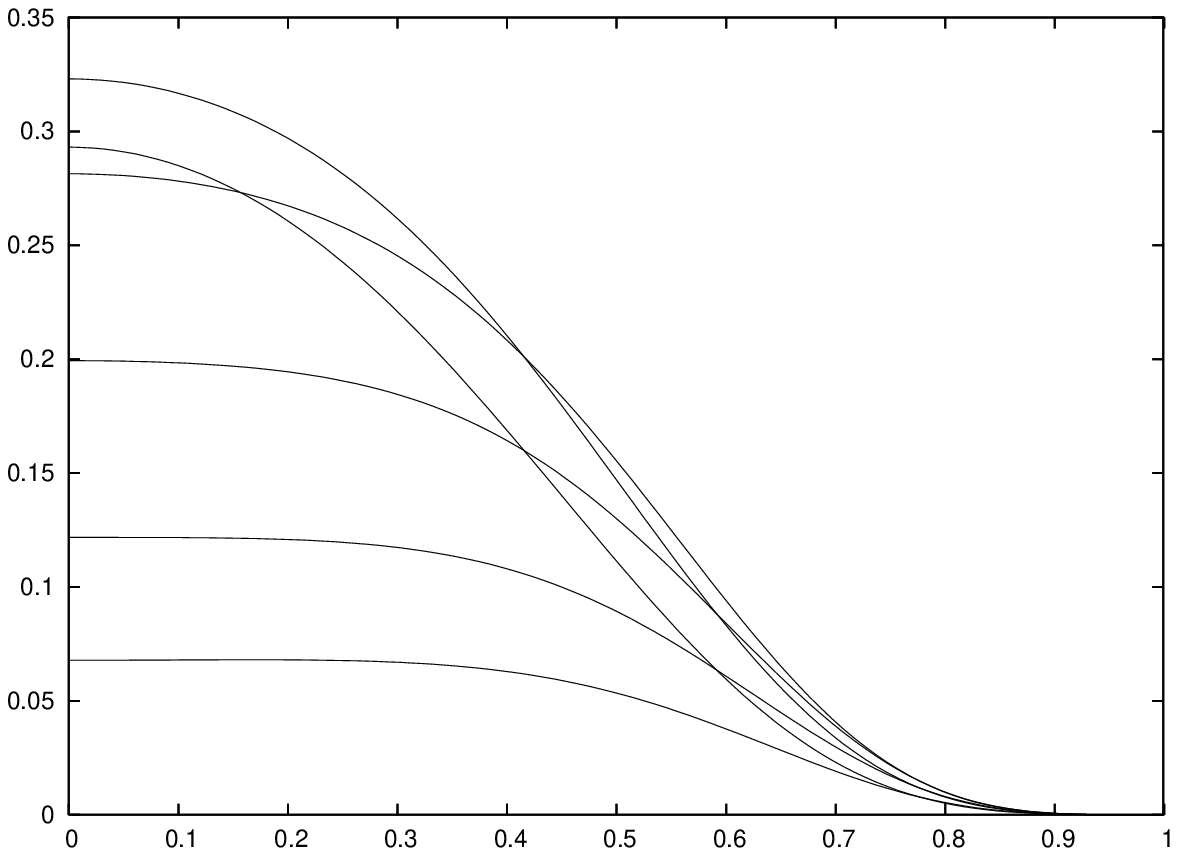} & 
\epsfig{width=2.75in,file=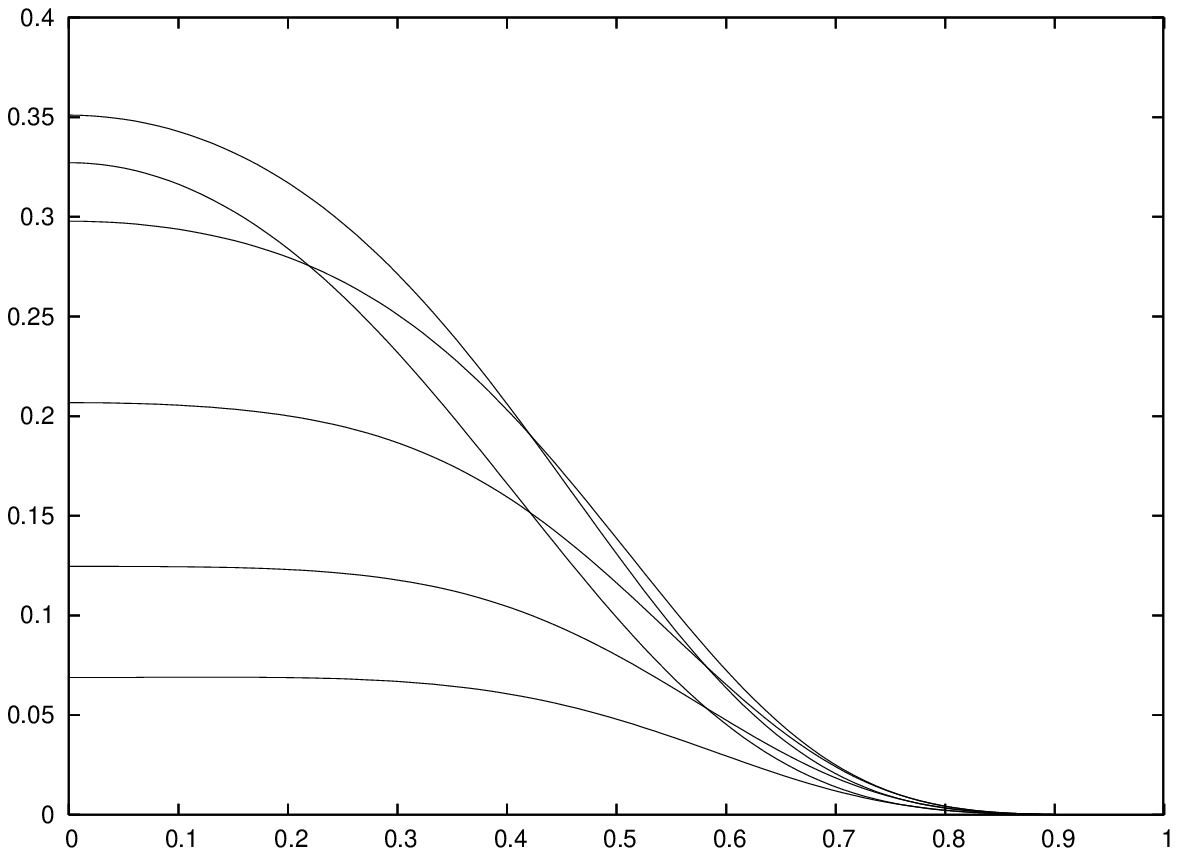} \\
 & \\
\end{array}$$
\caption{\label{figure1}Dimensionless surface energy density ${\tilde
\epsilon}_m$ as a function of ${\tilde r}$ for the first six disk models with $m
= 1, 2, 3, 4, 5, 6$. In each case, we plot ${\tilde \epsilon}_m ({\tilde r})$
for $0 \leq {\tilde r} \leq 1$ with different values of the parameter ${\tilde
k}$. We first take ${\tilde k} = - 0.25$, the bottom curve in each plot, and
then ${\tilde k} = - 0.5$, $- 1$, $- 2$, $- 4$ and $- 8$.}
\end{figure*}

We plot, in Figure \ref{figure1}, the dimensionless surface energy density
${\tilde \epsilon}_m$ as a function of ${\tilde r}$ for the first six disk
models with $m = 1$, $2$, $3$, $4$, $5$ and $6$. In each case, we plot ${\tilde
\epsilon}_m ({\tilde r})$ for $0 \leq {\tilde r} \leq 1$ with different values
of the parameter ${\tilde k}$. We first take ${\tilde k} = - 0.25$, the bottom
curve in each plot, and then ${\tilde k} = - 0.5$, $- 1$, $- 2$, $- 4$ and $-
8$. As we can see, in all the cases the energy density is everywhere positive
and vanishes at the edge of the disk. However, there are different behaviors
depending of the values of $m$ and ${\tilde k}$.  So, for the first two models,
for $m = 1$ and $m = 2$, we find that for small values of $|{\tilde k}|$ the
energy density presents a maximum near the edge of the disk, whereas that for
higher values of $|{\tilde k}|$ the maximum occurs at the center of the disk. On
the other hand, for $m > 3$, we find that for all the values of $|{\tilde k}|$
the maximum of the energy density occurs at the center of the disk.

\begin{figure}
$$\begin{array}{c}
{\tilde \epsilon}_m \\
\epsfig{width=2.75in,file=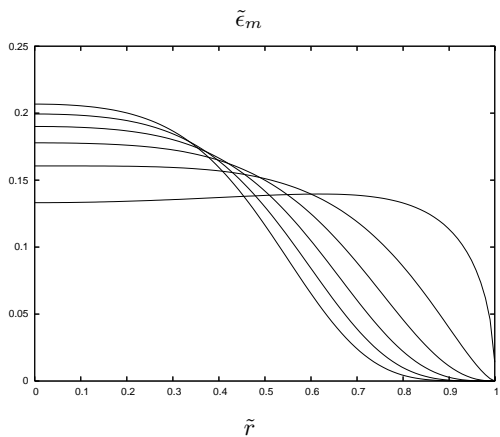} \\
{\tilde r} \\
\end{array}$$
\caption{\label{figure2} Dimensionless energy density ${\tilde \epsilon}_m$ as a
function of ${\tilde r}$ with ${\tilde k} = - 1$ for the six disk models with $m
= 1$, ... , $6$.}
\end{figure}

\begin{figure}
$$\begin{array}{c}
{\tilde \epsilon}_m (0) \\
\epsfig{width=2.75in,file=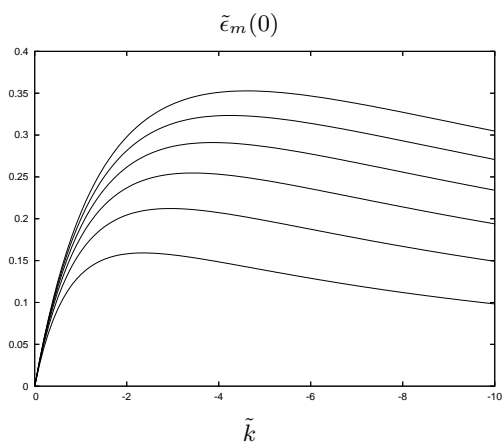} \\
{\tilde k} \\
\end{array}$$
\caption{\label{figure3}Dimensionless energy density at the center of the disks,
${\tilde \epsilon} (0)$, as a fuction of ${\tilde k}$ for the same six disk
models previously considered.}
\end{figure}

We can also see that as the value of $m$ increases, the energy density is more
concentrated at the center of the disks. Also, in the central part of the disks,
as $m$ increases the value of the energy density also increases in such a way
that, for a given value of ${\tilde k}$, the energy density is grater for a
greater value of $m$. On the other hand, at the border of the disks the behavior
is opposite, the value of the energy density decreases as $m$ increases. In
order to see this, we plot in Figure \ref{figure2} the dimensionless energy
density ${\tilde \epsilon}_m$ as a function of ${\tilde r}$ with ${\tilde k} = -
1$ for the disk models with $m = 1$, ... , $6$. Finally, for a given value of
$m$, the value of the energy density initially increases with the value of
$|{\tilde k}|$, but then it reaches a maximum and then it decreases as $|{\tilde
k}|$ increases. We depict this behavior at Figure \ref{figure3} where we plot
the dimensionless energy density at the center of the disks, ${\tilde \epsilon}
(0)$, as a fuction of ${\tilde k}$ for the same six disk models previously
considered. Now then, as the charge density of the disks is equal, up tu a sign, to their energy density, all the previous analysis apply as well to the behavior of the surface charge density of the models.


\section{\label{sec:conc}Concluding Remarks}

We presented an infinite family of axisymmetric charged dust disks of finite
extension with well behaved surface energy and charge densities. The disk were
obtained by solving the vacuum Einstein-Maxwell equations system for a
conformastatic spacetime. In order to obtain the solutions, a functional
dependence was assumed between the electric potential and the metric function
and beteween this one and an auxiliary function. The solutions were then
expressed in terms of a solution of Laplace equation corresponding to a family
of Newtonian thin disks of finite radius, the generalized Kalnajs disks
\cite{GR}, which describes a well behaved family of flat galaxy models.

The relativistic thin disks here presented have a charge density that is equal,
up to a sign, to their energy density, and so they are examples of the commonly
named `electrically counterpoised dust' equilibrium configuration. The energy
density of the disks is everywhere positive and well behaved, vanishing at the
edge. Also, as the value of $m$ increases, the energy density is more
concentrated at the center of the disks, having a maximum at $r = 0$ for all the
values of $|{\tilde k}|$. However, for the first two models, for $m = 1$ and $m
= 2$, for small values of $|{\tilde k}|$ the energy density presents a maximum
near the edge of the disk, whereas that for higher values of $|{\tilde k}|$ the
maximum occurs at the center of the disk.

Furthermore, as the energy density of the disks is everywhere positive and the
disks are made of dust, all the models are in a complete agreement with all the
energy conditions, a fact of particular relevance in the study of relativistic
thin disks models. Indeed, as was mentioned at the introduction, many of the
relativistic thin disks models that had been studied in the literature do not
fully agrees with these conditions.

Now then, as we can see from the equations at Sec. \ref{sec:ecs}, the procedure
here presented can be applied not only for axisymmetric conformastatic
spacetimes but can also be used to obtain non-axisymmetric solutions of the
vacuum Einstein-Maxwell equations. Accordingly, the thin disks models here
presented can be generalized by considering for the auxiliary function $U$
solutions of Laplace equations without the impossed axial symmetry. So, we are
now working in this direction and the results will be presented in a next paper.
Analogously, the generalization to conformastationary spacetimes with magnetic
fields is in consideration.

\begin{acknowledgments}

A. C. G-P. wants to thank the financial support from COLCIENCIAS, Colombia.
Also, A. C. G-P would like to thank to T. Ledvinka by the very helpful
discussions.

\end{acknowledgments}

\end{document}